\documentclass[useAMS,usenatbib]{mn2e}

\usepackage{natbib}
\usepackage{graphicx}
\usepackage{amssymb}
\usepackage{color}
\usepackage{caption}
\usepackage{lipsum,graphicx,multicol}
\usepackage{float}

\usepackage{subcaption}

\title[AGN outflows: models vs. observations]{AGN-driven galactic outflows: comparing models to observations}
\author[ ]
{W. Ishibashi$^{1}$\thanks{E-mail: wako.ishibashi@physik.uzh.ch}, A. C. Fabian$^{2}$ and N. Arakawa$^{2}$
\footnotemark[0]\\
\footnotemark[0]\\
$^{1}$Physik-Institut, Universitat Zurich, Winterthurerstrasse 190, 8057 Zurich, Switzerland \\
$^{2}$Institute of Astronomy, Madingley Road, Cambridge CB3 0HA 
}

\begin{document}

\pdfminorversion=4

\date{Accepted ? Received ?; in original form ? }

\pagerange{\pageref{firstpage}--\pageref{lastpage}} \pubyear{2012}

\maketitle

\label{firstpage}

\begin{abstract} 
The actual mechanism(s) powering galactic outflows in active galactic nuclei (AGN) is still a matter of debate. At least two physical models have been considered in the literature: wind shocks and radiation pressure on dust. Here we provide a first quantitative comparison of the AGN radiative feedback scenario with observations of galactic outflows. We directly compare our radiation pressure-driven shell models with the observational data from the most recent compilation of molecular outflows on galactic scales. We show that the observed dynamics and energetics of galactic outflows can be reproduced by AGN radiative feedback, with the inclusion of radiation trapping and/or luminosity evolution. The predicted scalings of the outflow energetics with AGN luminosity can also quantitatively account for the observational scaling relations. Furthermore, sources with both ultra-fast and molecular outflow detections are found to be located in the `forbidden' region of the $N_\mathrm{H} - \lambda$ plane. Overall, an encouraging agreement is obtained over a wide range of AGN and host galaxy parameters. We discuss our results in the context of recent observational findings and numerical simulations. In conclusion, AGN radiative feedback is a promising mechanism for driving galactic outflows that should be considered, alongside wind feedback, in the interpretation of future observational data.  
\end{abstract} 



\section{Introduction}
\label{Section_Introduction}

Outflows on galactic scales are now commonly observed in active galactic nuclei (AGN) host galaxies. The outflows are detected in ionised, neutral, and molecular gas phases, indicative of a multi-phase nature \citep[e.g.][and references therein]{Sturm_et_2011, Cicone_et_2014, Fiore_et_2017, Fluetsch_et_2019, Veilleux_et_2020}. Among the different gas phases, the cool molecular gas is the dominant component carrying the bulk of the outflowing mass. In fact, the molecular phase may account for the majority of the total mass outflow rate in galaxies, with a greater fraction (up to $\sim 90 \%$) in AGN-dominated systems \citep{Fluetsch_et_2020}. As the molecular gas is also the primary fuel for star formation, molecular outflows have important implications for galaxy evolution. 

Observations of galactic molecular outflows indicate high velocities (up to $\sim 1000$ km/s) on kpc-scales, associated with large momentum rates (several times the photon momentum $L/c$) and high kinetic powers (a few percent of the AGN luminosity) \citep[e.g.][]{Fluetsch_et_2019}. Despite the remarkable observational progress in quantifying the dynamics and energetics of galactic outflows, their physical origin remains unclear. There is ongoing debate trying to determine whether outflows are powered by jets, winds, or radiation. In this context, two main physical models have been proposed and discussed in the literature:  wind shocks and radiation pressure on dust. Radio jets can also drive galactic outflows, but this feedback mode may be mostly limited to radio-loud sources, which only form a minority of the total AGN population. In addition, cosmic rays may help the launching of cool dense outflows with a smooth gas distribution, as observed in magneto-hydrodynamical simulations including cosmic ray pressure \citep[e.g.][]{Girichidis_et_2018}. The potential role of cosmic ray-driven outflows in galaxy evolution is now gaining interest \citep[see][and references therein]{Veilleux_et_2020} but will not be further pursued here. 

In the `wind feedback' model, a fast nuclear wind (or ultra-fast outflow UFO) shocks against the cold interstellar medium of the host galaxy, leading to wind energy-driven outflows propagating on large scales \citep[][and references therein]{Zubovas_King_2012, King_Pounds_2015}. The wind energy-driving mode predicts large momentum boosts ($\dot{p} \sim 20 \, L/c$) and high kinetic powers ($\dot{E}_k \sim 0.05 L$) for fiducial parameters. These numbers are consistent with observational measurements, and the wind energy-driven mechanism has been the favoured interpretation for the high energetics observed in galactic outflows. A recent work extends the wind shock model to a two-dimensional configuration in a disc geometry and shows that the shock expansion tends to follow the path of least resistance \citep{Menci_et_2019}. Based on a detailed comparison between model outputs and outflow data, an overall good agreement is reported. 

In the `radiation feedback' scenario, galactic outflows are driven by radiation pressure on dusty gas, due to the enhanced radiation-matter coupling \citep{Fabian_1999, Murray_et_2005, Thompson_et_2015}. We note that radiation pressure-driving has often been disregarded on the grounds that it is unable to yield large momentum boosts. This may be the case in the single scattering limit, but the outcome is different when one takes into account the trapping of reprocessed radiation. In fact, in contrast to common belief, radiation pressure-driven outflows can also attain large values of the momentum rates ($\dot{p} \gtrsim 10 L/c$) and kinetic powers ($\dot{E}_k \sim \mathrm{few} \%$ L), provided that radiation trapping is properly included \citep{Thompson_et_2015, Ishibashi_Fabian_2015, Ishibashi_et_2018a, Costa_et_2018}. Therefore radiation feedback should be considered as a viable mechanism for driving galactic outflows, alongside wind feedback. 

Up to now, galactic outflow observations have been preferentially interpreted in the framework of the wind energy-driven outflow model. However, some recent observational works start to question wind feedback as the only viable and most favoured mechanism. 
In a number of sources hosting both a small-scale UFO and a large-scale molecular outflow, the measured momentum boosts are found to be significantly lower than the values predicted by a purely wind energy-conserving scenario \citep[e.g.][]{Bischetti_et_2019, Sirressi_et_2019, Reeves_Braito_2019}. This suggests the possibility that other mechanisms, such as radiation pressure on dust, may play a role in driving the large-scale galactic outflows. The relative importance of the hot wind and radiation pressure can also be empirically constrained by emission line diagnosis. The line ratios observed in quasar outflows suggest that radiation pressure can dominate hot gas pressure over a wide range of radii \citep{Stern_et_2016, Somalwar_et_2020}.

It is thus timely to reconsider AGN radiative feedback as an alternative mechanism for powering galactic outflows. Here we provide a first quantitative comparison of the radiation pressure-driven outflow model with observations of galactic molecular outflows. The paper is structured as follows. We start by recalling the basics of AGN feedback driven by radiation pressure on dust (Section \ref{Section_Model}). In Section \ref{Section_Comparison}, we first provide a detailed individual comparison with a prototype source (Sect. \ref{Section_Individual_Comparison}), and then compare the radiation pressure-driven shell models with a large sample of molecular outflows compiled from the literature (Sect. \ref{Section_Global_Comparison}). We consider the resulting outflow scaling relations in Section \ref{Section_Scaling_Relations}, and analyse the location of the outflowing sources in the $N_\mathrm{H} - \lambda$ plane (Section \ref{Section_NH_lambda}). We discuss the global results in relation to other physical models and recent observational findings (Section \ref{Section_Discussion}) and conclude in Section \ref{Section_Conclusion}. 


\section{The radiation feedback model}
\label{Section_Model}

We briefly recall the basics of AGN radiative feedback due to radiation pressure on dust \citep[see our previous papers for more details, e.g.][]{Ishibashi_Fabian_2015, Ishibashi_et_2018a}. 
The equation of motion of the radiation pressure-driven shell is given by
\begin{equation}
\frac{d}{dt} [M_\mathrm{sh} v] = \frac{L}{c} (1 + \tau_\mathrm{IR} - e^{-\tau_\mathrm{UV}} ) - \frac{G M(r) M_\mathrm{sh}}{r^2} , 
\label{Eq_motion}
\end{equation}
where $L$ is the central luminosity, $M_\mathrm{sh}$ is the shell mass, and $M(r) = \frac{2 \sigma^2 r}{G}$ is the total mass distribution assuming an isothermal potential with velocity dispersion $\sigma$. 
The infrared (IR) and ultraviolet (UV) optical depths are respectively given by
\begin{equation}
\tau_\mathrm{IR} = \frac{\kappa_\mathrm{IR} M_\mathrm{sh}}{4 \pi r^2} ,
\end{equation}
\begin{equation}
\tau_\mathrm{UV} = \frac{\kappa_\mathrm{UV} M_\mathrm{sh}}{4 \pi r^2} , 
\end{equation}
where $\kappa_\mathrm{IR}$=$5 \, \mathrm{cm^2 g^{-1} f_{dg, MW}}$ and $\kappa_\mathrm{UV}$=$10^3 \, \mathrm{cm^2 g^{-1} f_{dg, MW}}$ are the IR and UV opacities, with the dust-to-gas ratio normalised to the Milky Way value.

In order to launch an outflow, a certain minimal luminosity needs to be exceeded. 
The critical luminosity is obtained by equating the radiative force to the gravitational force:
\begin{equation}
L_c = \frac{2 c \sigma^2 M_\mathrm{sh}}{r} (1 + \tau_\mathrm{IR} - e^{-\tau_\mathrm{UV}} )^{-1} , 
\label{Eq_Lc}
\end{equation}
which may be considered as a generalised form of the effective Eddington luminosity. 

Integrating the equation of motion (Eq. \ref{Eq_motion}), and neglecting the UV exponential term, we obtain the analytic solution for the velocity of the outflowing shell
\begin{equation}
v \cong \sqrt{ \frac{2 L}{c M_\mathrm{sh}} (r - R_0) + \frac{\kappa_\mathrm{IR} L}{2 \pi c} ( \frac{1}{R_0} - \frac{1}{r} ) } \, , 
\label{Eq_v_analytic}
\end{equation}
where $R_0$ is the initial radius and the logarithmic factor is ignored. 

By analogy with the observational works on galactic outflows, we define three model parameters quantifying the outflow energetics: the mass outflow rate ($\dot{M}$), the momentum rate ($\dot{p}$), and the kinetic power ($\dot{E}_k$):
\begin{equation}
\dot{M} = \frac{M_\mathrm{sh}}{t_\mathrm{flow}} 
= \frac{M_\mathrm{sh} v}{r} , 
\label{Eq_Mdot}
\end{equation} 
\begin{equation}
\dot{p}  = \dot{M} v 
= \frac{M_\mathrm{sh} v^2}{r} , 
\label{Eq_pdot}
\end{equation} 
\begin{equation}
\dot{E}_k = \frac{1}{2} \dot{M} v^2 
= \frac{M_\mathrm{sh} v^3}{2 r} .
\label{Eq_Ekdot}
\end{equation}
The above outflow energetics are computed based on the so-called thin-shell approximation adopted in many observational studies \citep[e.g.][]{Gonzalez-Alfonso_et_2017}. 

From the approximate form of the radial velocity profile (Eq. \ref{Eq_v_analytic}), we can derive the analytic limits for the mass outflow rate
\begin{equation}
\dot{M} 
= \left( \frac{2 L M_\mathrm{sh}}{c r^2} (r - R_0) + \frac{\kappa_\mathrm{IR} L M_\mathrm{sh}^2}{2 \pi c r^2} ( \frac{1}{R_0} - \frac{1}{r} ) \right)^{1/2} , 
\label{Eq_Mdot}
\end{equation} 
and kinetic power
\begin{equation}
\dot{E}_k = 
\frac{M_\mathrm{sh}}{2 r} \left( \frac{2 L}{c M_\mathrm{sh}} (r - R_0) + \frac{\kappa_\mathrm{IR} L}{2 \pi c} ( \frac{1}{R_0} - \frac{1}{r} ) \right)^{3/2} .
\label{Eq_Edotk}
\end{equation}

As a result, we obtain that the mass outflow rate scales with luminosity as 
\begin{equation}
\dot{M} \propto L^{1/2} , 
\end{equation}
while the kinetic power scales with luminosity as
\begin{equation}
\dot{E}_k \propto L^{3/2} ,
\end{equation}
implying a sub-linear scaling for the mass outflow rate and a super-linear scaling for the kinetic power \citep[see also][]{Ishibashi_et_2018a}. 

Moreover, two derived quantities are generally used in characterising the outflow energetics: the momentum ratio ($\zeta$) and energy ratio ($\epsilon_k$), defined by
\begin{equation}
\zeta = \frac{\dot{p}}{L/c} = M_\mathrm{sh} \frac{v^2}{r} \frac{c}{L} , 
\label{momentum_ratio}
\end{equation} 
\begin{equation}
\epsilon_k = \frac{\dot{E}_k}{L}  = \frac{1}{2} M_\mathrm{sh} \frac{v^3}{r} \frac{1}{L} .
\label{energy_ratio}
\end{equation}

In the radiative feedback scenario, large values of the outflow energetics are reached at small radii, where the outflowing shell is optically thick to the reprocessed IR radiation. In our picture, the maximal values of both the momentum ratio and energy ratio are mainly determined by the initial IR optical depth: $\zeta_\mathrm{IR} \sim 2 \sqrt{\tau_\mathrm{IR,0}}$ and $\epsilon_\mathrm{k,IR} \sim \sqrt{\tau_\mathrm{IR,0}} \frac{v_\mathrm{IR}}{c}$ (where $v_\mathrm{IR}$ is the velocity near the IR transparency radius). 
We have previously discussed how AGN radiative feedback may qualitatively explain the observed outflow energetics, and analysed the effects of luminosity evolution and shell mass configurations \citep{Ishibashi_et_2018a, Ishibashi_Fabian_2018}. 


\section{Comparison with observations: outflow energetics}
\label{Section_Comparison}

Here we provide a quantitative comparison of the AGN radiative feedback model (outlined in Section \ref{Section_Model}) with observational data of galactic molecular outflows. 


\subsection{Individual comparison}
\label{Section_Individual_Comparison}

\begin{figure}
\begin{center}
\includegraphics[angle=0,width=0.4\textwidth]{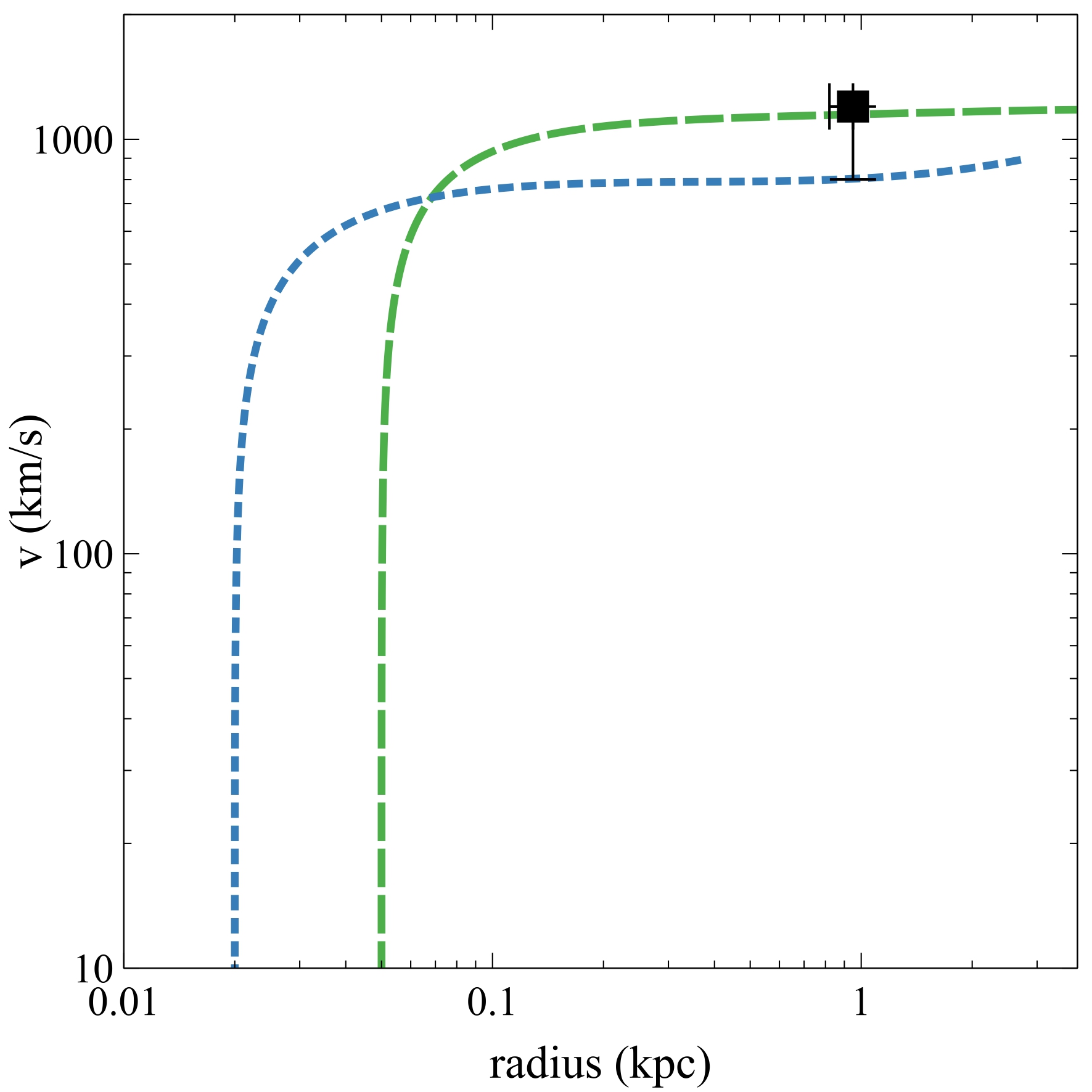} 
\caption{ 
Comparison of the radial velocity profile of radiation pressure-driven outflows with observations of IRAS F08572+3915. 
Constant luminosity case: $L \sim 4 \times 10^{45}$ erg/s, $R_0 = 20$ pc, $f_\mathrm{dg} = 5 \times f_\mathrm{dg,MW}$ (blue dotted). Luminosity decay case: $L(t) = L_0 (1+ t/t_d)^{-\delta}$ with $L_0 = 4 \times 10^{46}$ erg/s, $t_d = 10^5$ yr, $\delta = 1$, $R_0 = 50$ pc, $f_\mathrm{dg} = 3.5 \times f_\mathrm{dg,MW}$ (green dashed). 
The observational data point (black square) is from the latest CO observations \citep{Herrera-Camus_et_2020}, and the vertical/horizontal bars indicate previous measurements \citep{Cicone_et_2014, Fluetsch_et_2019}. 
}
\label{Fig_IRAS_velocity}
\end{center}
\end{figure} 

\begin{figure*}
\begin{multicols}{3}
    \includegraphics[width=\linewidth]{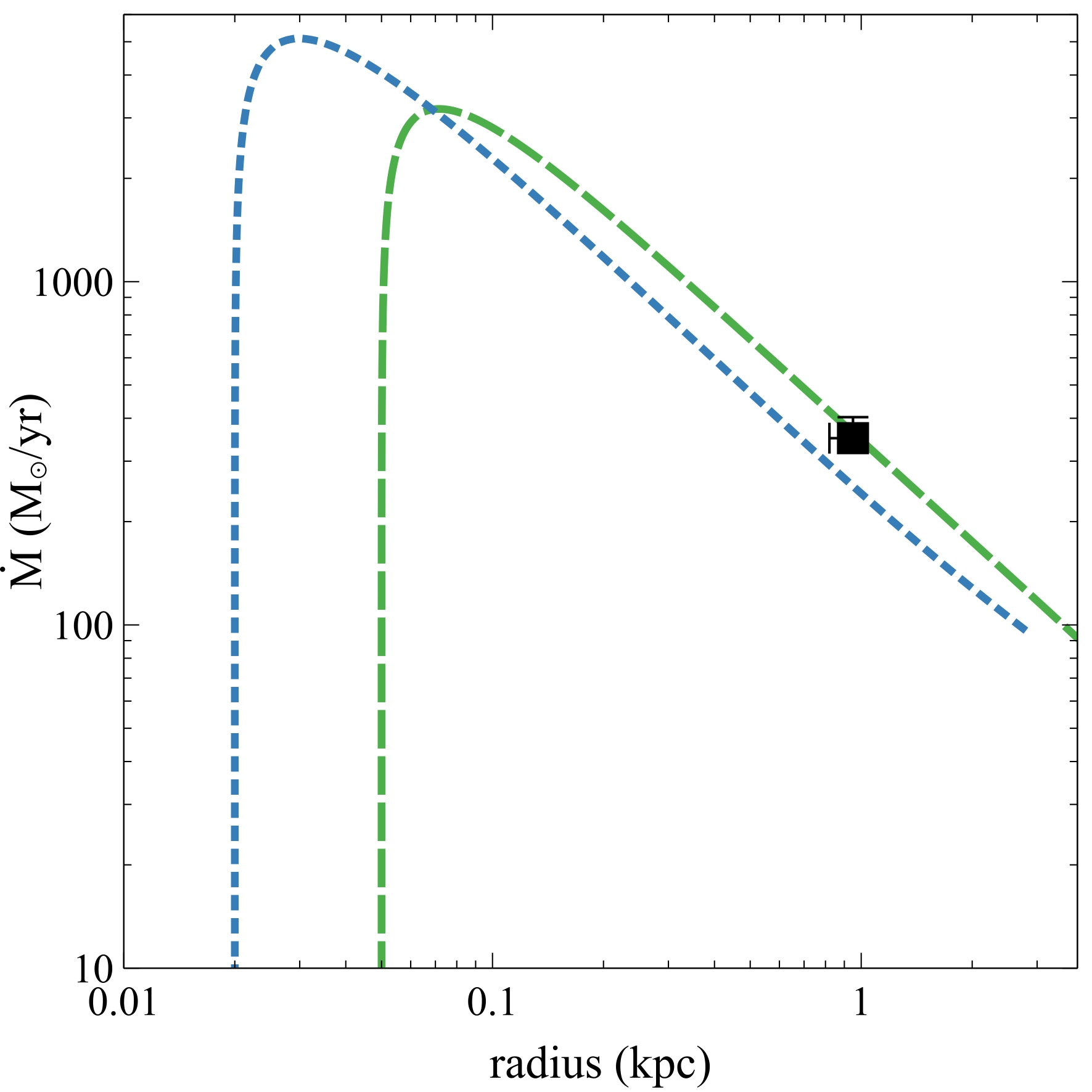}\par 
    \includegraphics[width=\linewidth]{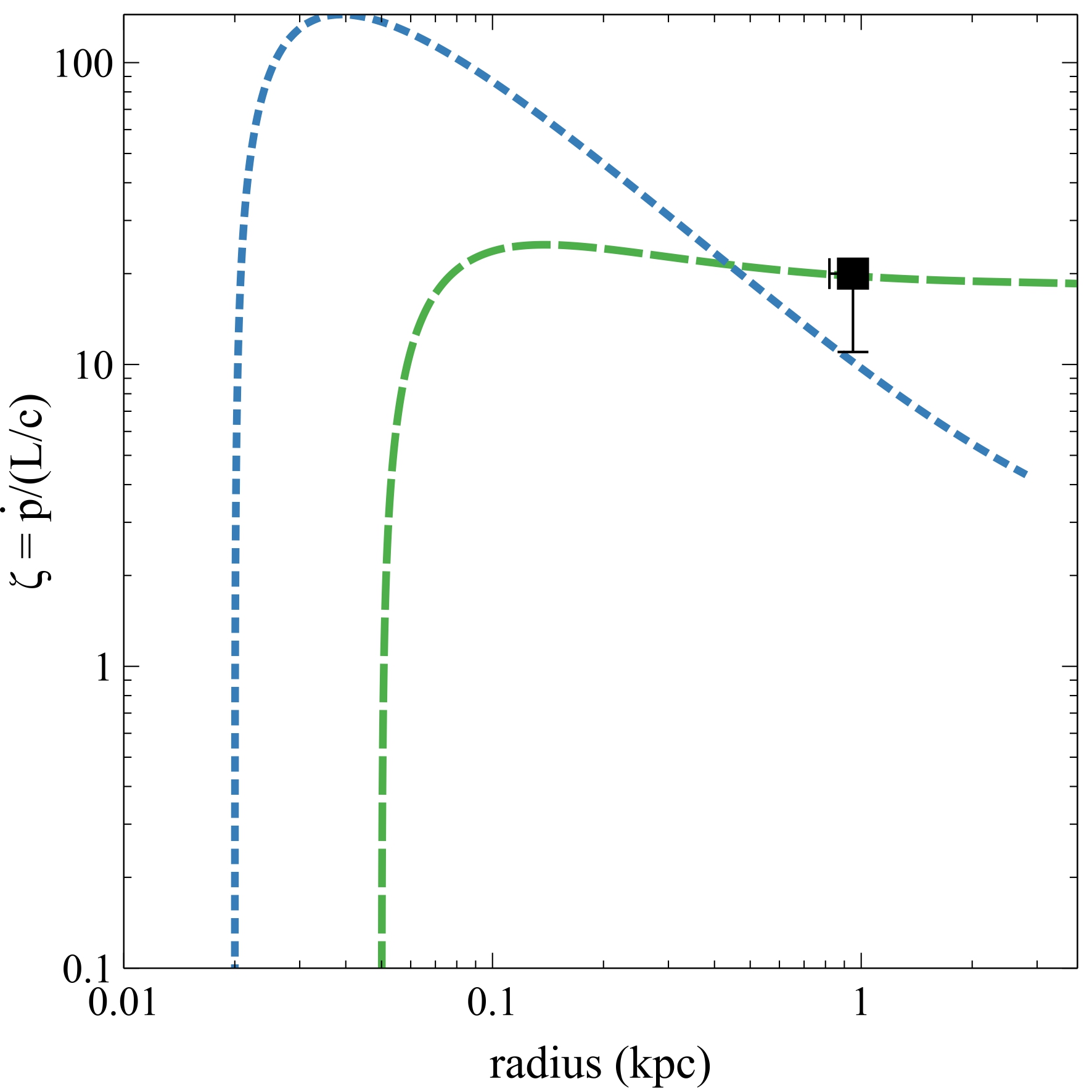}\par 
    \includegraphics[width=\linewidth]{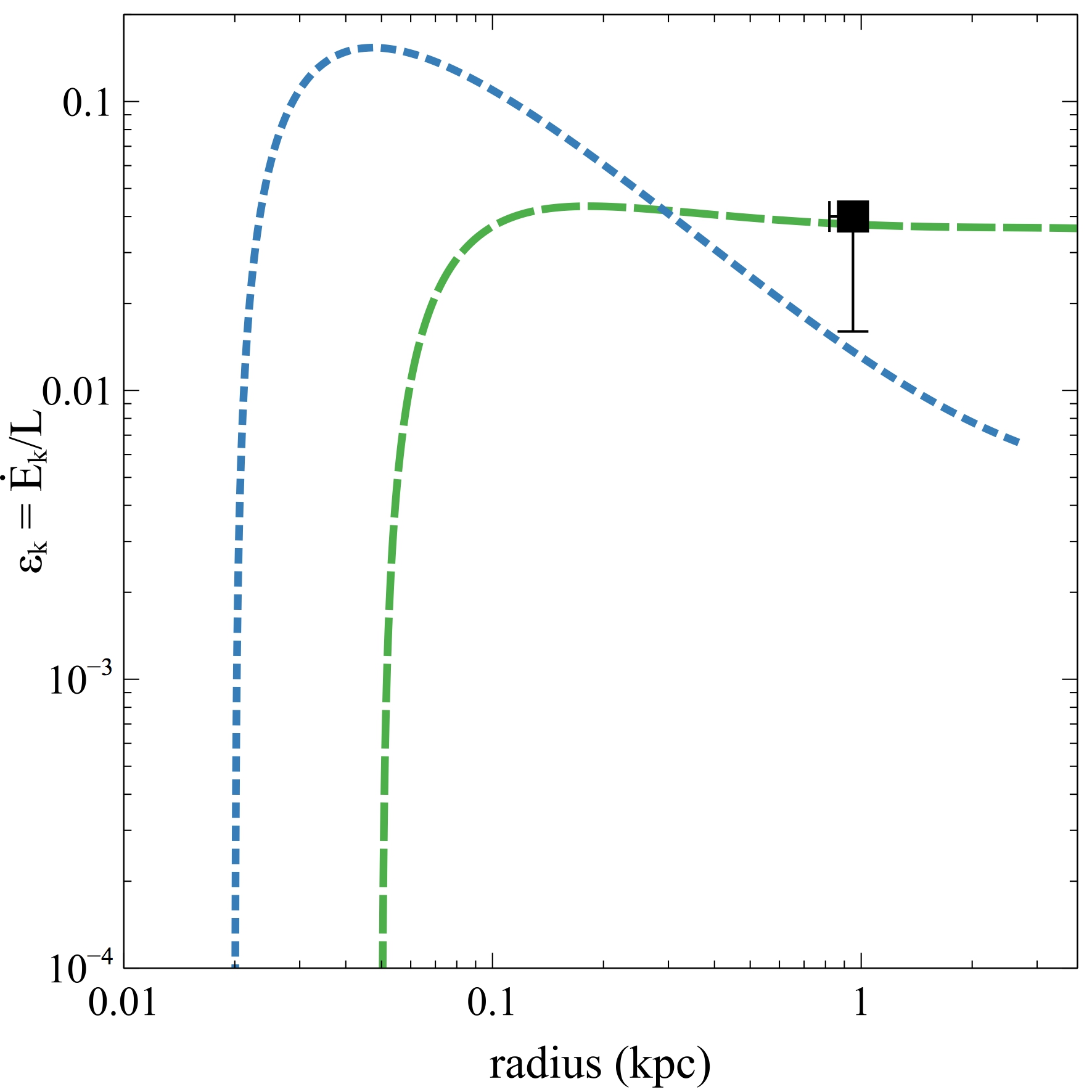}\par
    \end{multicols}
\caption{ 
Comparison of the outflow energetics of radiation pressure-driven outflows with observations of IRAS F08572+3915. 
Radial profiles of the mass outflow rate $\dot{M}$ (left panel), momentum ratio $\zeta$ (middle panel), and energy ratio $\epsilon_k$ (right panel). The represented model cases and observational data are the same as in Figure \ref{Fig_IRAS_velocity}. 
}
\label{Fig_IRAS_energetics}
\end{figure*}

We first compare the radiation pressure-driven outflow model with observations of a particular individual source: IRAS F08572+3915. This system is a key example of a local dust-obscured ultra-luminous infrared galaxy (ULIRG), with AGN luminosity of $L \sim 4 \times 10^{45}$ erg/s. A molecular outflow was previously detected in CO observations, with an average velocity of $\sim 800$ km/s \citep{Cicone_et_2014}, and the corresponding outflow energetics are reported in \citet{Fluetsch_et_2019}. 
New high angular resolution CO observations spatially resolve the molecular outflow, which has an outflowing mass of $M \sim 3 \times 10^8 M_{\odot}$ and reaches a velocity of $v \sim 1200$ km/s around $\sim 1$ kpc \citep{Herrera-Camus_et_2020}. 
The associated mass outflow rate is $\dot{M} \sim 350 \, \mathrm{M_{\odot}/yr}$, while the outflow momentum rate and kinetic power are $\dot{p} \sim 3 \times 10^{36}$ dyn and $\dot{E}_k \sim 2 \times 10^{44}$ erg/s, respectively. The corresponding momentum ratio is of order $\zeta = \dot{p}/(L/c) \sim 20$ and the energy ratio is $\epsilon_k = \dot{E}_k/L \sim 0.04$. The high values of the outflow energetics measured in this source have been exclusively interpreted in the framework of the wind energy-driving scenario \citep{Herrera-Camus_et_2020}. 

Here we examine whether the outflow energetics observed in IRAS F08572+3915 can be explained in terms of radiation pressure on dust. The available observational values of the source \citep{Herrera-Camus_et_2020} are adopted as input parameters of the radiation-driven outflow model. We follow the evolution of an outflowing mass of $M_\mathrm{sh} = 3 \times 10^8 M_{\odot}$, launched from initial radii of $R_0 = (20-50)$ pc, propagating in an isothermal potential with velocity dispersion $\sigma = 150$ km/s. Assuming a Milky Way-like dust-to-gas ratio, we find that a large-scale massive outflow can not be powered by the present-day luminosity. 
We therefore consider two possibilities that may help enhance the outflow powering: IR radiation trapping and AGN luminosity evolution \citep{Ishibashi_Fabian_2018}. 

In the nuclear regions of dense starbursts and obscured AGNs, large gas masses ($\sim 10^8 M_{\odot}$) are likely concentrated in the inner tens of parsec scales. Millimetre observations reveal the existence of compact ($r <$ 15-75 pc) obscured nuclei characterised by very high column densities \citep[][and references therein]{Aalto_et_2019b}. Indeed, Compton-thick obscuration may be expected in compact obscured nuclei, as suggested by recent observations (see Section \ref{Section_Discussion}). 
Moreover, large amounts of dust are likely expected in the nuclei of ULIRG-like systems, yielding high dust-to-gas ratios \citep[e.g.][]{Gowardhan_et_2018}. These peculiar environments should form favourable conditions for AGN radiative feedback. 

Another possibility is AGN luminosity evolution. It is established that AGNs are intrinsically variable objects, which can undergo luminosity variations of several orders of magnitude over a range of timescales. In fact, it is rather unlikely that the central luminosity stayed constant during the whole time span required for the outflow to reach the current location (a typical crossing time may be of order $t \sim r/v \sim 10^6$ yr for $r \sim 1$ kpc and $v \sim 1000$ km/s). A power-law decay in AGN luminosity may be expected, for instance if the accretion disc slowly dissipates on a viscous timescale \citep{King_Pringle_2007}. Luminosity variations with power-law decay forms have also been considered in the framework of the wind shock model \citep{Zubovas_2018}.  

In Figure \ref{Fig_IRAS_velocity}, we plot the radial velocity profile of the radiation pressure-driven shells compared with the observational data of IRAS F08572+3915. The data point from the latest CO observations (black square) is plotted along with previous measurements (vertical/horizontal bars) representing a plausible range. For a constant luminosity equal to the present-day value ($L \sim 4 \times 10^{45}$ erg/s), and enhanced dust-to-gas ratio ($f_\mathrm{dg} = 5 \times f_\mathrm{dg,MW}$), a high-velocity outflow can propagate on galactic scales (blue dotted line). But this requires quite extreme initial conditions (with a column density of $N_0 \sim 8 \times 10^{24} \mathrm{cm^{-2}}$ and initial IR optical depth of $\tau_\mathrm{IR,0} \sim 330$). Alternatively, there is the possibility that the central luminosity was higher in the past, due to the variable nature of the AGN. Assuming a luminosity evolution of the form $L(t) = L_0 (1 + t/t_d)^{-1}$ (power-law decay) with an initial luminosity of $L_0 = 4 \times 10^{46}$ erg/s and characteristic timescale $t_d = 10^5$yr, the resulting outflow can be accelerated to velocities of $v \sim 1000$ km/s on kpc-scales (green dashed line). In this case, the required initial conditions can be more realistic, with an initial column density of $N_0 \sim 10^{24} \mathrm{cm^{-2}}$ and $\tau_\mathrm{IR,0} \lesssim 40$. 

In Figure \ref{Fig_IRAS_energetics}, we show the corresponding outflow energetics: the mass outflow rate $\dot{M}$ (left panel), the momentum ratio $\zeta = \dot{p}/(L/c)$ (middle panel) and the energy ratio $\epsilon_k = \dot{E}_k/L$ (right panel) as a function of radius. We observe that large values of the momentum ratio ($\zeta \gtrsim 10$) and energy ratio ($\epsilon_k \sim \textrm{a few} \%$) can be accounted for, provided that the outflow is initially optically thick to the reprocessed IR radiation. 
In particular, the IRAS F08572+3915 data can be adequately reproduced by considering a radiatively-driven outflow coupled with AGN luminosity decay. Such a luminosity evolution may be expected in the time span between the outflow launch and its current location, while large IR optical depths are likely reached in the buried nuclei of ULIRG-like systems. 


\subsection{Global comparison}
\label{Section_Global_Comparison}

We next compare the radiation pressure-driven outflow model with the most recent compilation of CO molecular outflows in the local Universe \citep{Fluetsch_et_2019}. We consider all objects identified as AGNs (Seyferts and LINERs), but exclude sources classified as `fossil' outflow candidates. We also include two local molecular outflows recently reported in the literature, MCG-03-58-007 \citep{Sirressi_et_2019} and PDS 456 \citep{Bischetti_et_2019}. We further complement the local sample with the addition of four molecular outflows detected at higher redshifts ($z > 1.5$): HS 0810+2554 \citep{Chartas_et_2020}, XID 2028 \citep{Brusa_et_2018}, zC400528 \citep{Herrera-Camus_et_2019}, and APM 08279+5255 \citep{Feruglio_et_2017}. The total sample consists of 28 sources with CO molecular outflow measurements, and covers a wide range of AGN luminosities ($L \sim 10^{42}-10^{47}$ erg/s) and outflowing gas mass ($M_\mathrm{out} \sim 10^6-10^9 M_{\odot})$. The outflow energetics are all computed assuming the thin-shell approximation (cf. Section \ref{Section_Model}).  

\begin{figure*}
\begin{multicols}{2}
    \includegraphics[width=0.8\linewidth]{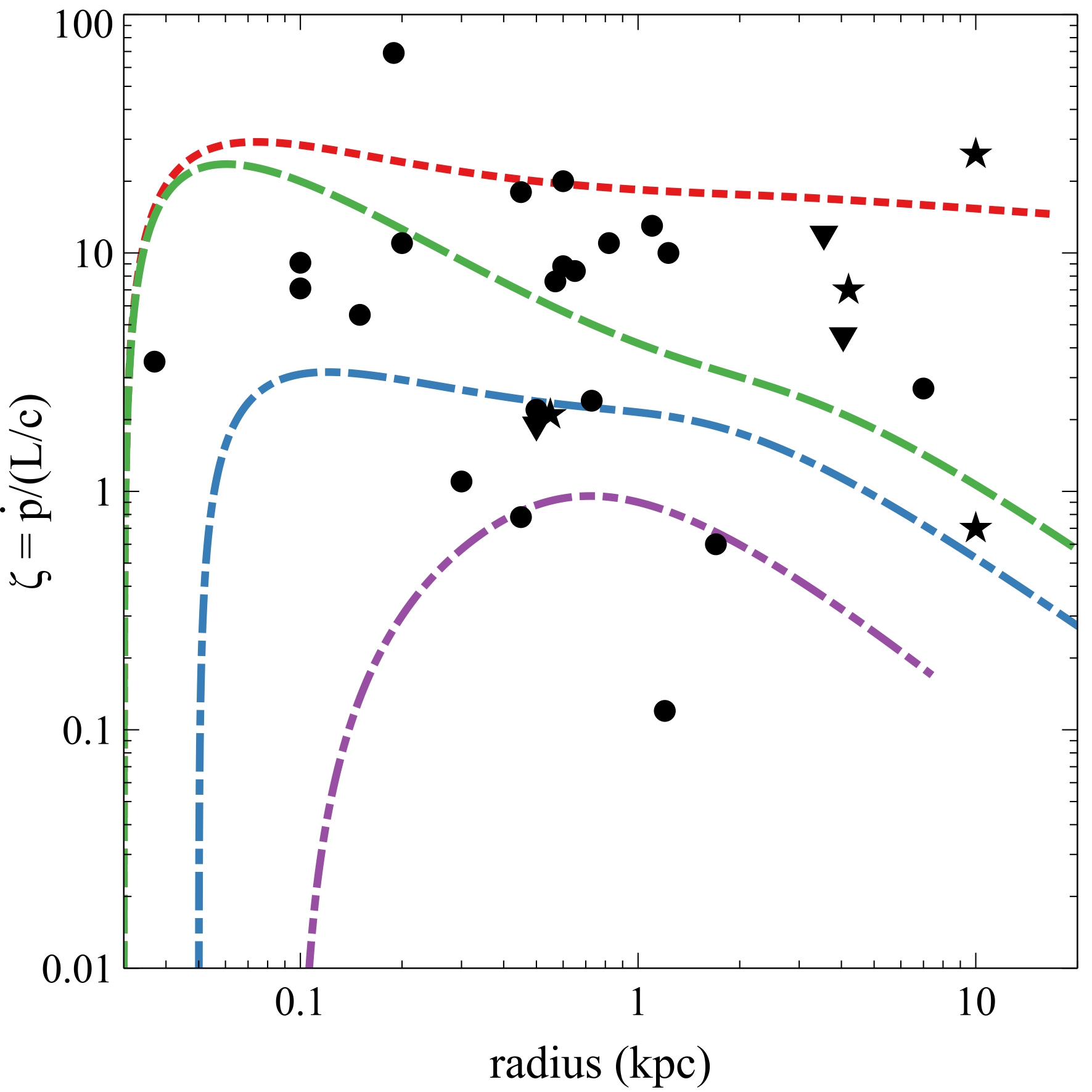}\par
    \includegraphics[width=0.8\linewidth]{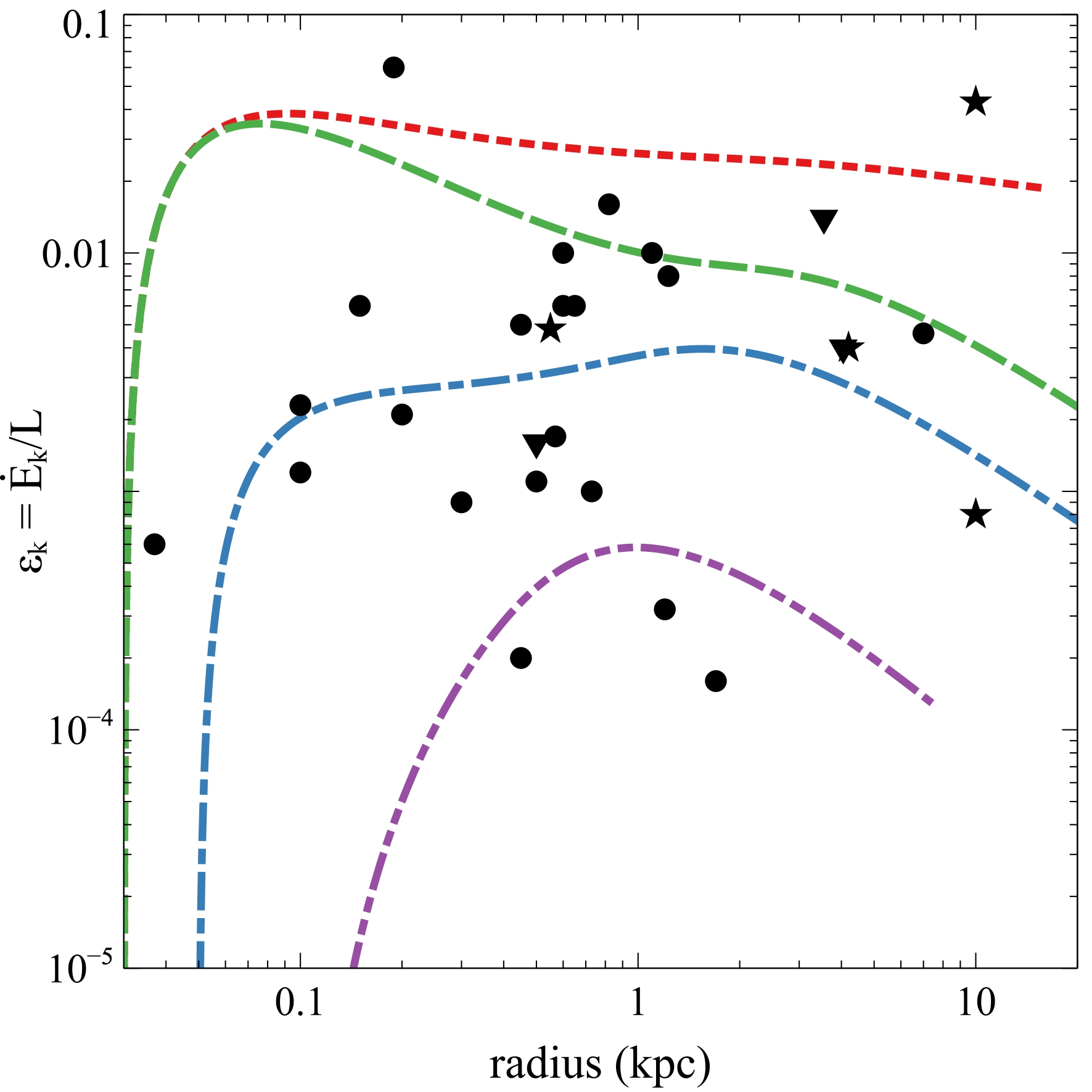}\par 
    \end{multicols}
\caption{
Comparison of the outflow energetics observed in a sample of molecular outflows (black symbols) with radiation pressure-driven outflow models (coloured curves). 
Observational samples: local molecular outflows (dots), upper limits (triangles), additional high-z sources (stars). Model cases: $L = 10^{45}$ erg/s, $\tau_\mathrm{IR,0} \sim 0.3$ (violet dash-dot-dot); $L = 5 \times 10^{45}$ erg/s, $\tau_\mathrm{IR,0} \sim 5$ (blue dash-dot); $L = 10^{46}$ erg/s, $\tau_\mathrm{IR,0} \sim 50$ (green dashed); $L(t) = L_0 \left( 1 + t/t_d \right)^{-\delta}$, with $L_0 = 10^{46}$ erg/s, $\delta = 1$, $t_d = 10^5$ yr  (red dotted). 
} 
\label{plot_energetics_sample}
\end{figure*}

In Figure \ref{plot_energetics_sample}, we plot the measurements of the momentum ratio and energy ratio of the molecular outflows at the observed radial location (denoted by different black symbols). In the same plots, we show the corresponding radial profiles of the energetics of radiation pressure-driven outflows (coloured curves). The different model curves represent different physical conditions in the sources, ranging from IR-optically thin cases to heavily obscured systems (see the caption of Fig. \ref{plot_energetics_sample} for numerical values). Compared to previous observational works, the latest outflow compilation covers a broader range in the measured outflow properties with a larger scatter, likely due to less biased samples \citep[as noted in][]{Fluetsch_et_2019}. 

From the left panel of Figure \ref{plot_energetics_sample}, we see that the molecular outflows span a range of momentum boosts, with the majority having values in the range $\zeta \sim 1-20$. In the radiative feedback scenario, low momentum ratios ($\zeta \sim 1$) are easily obtained for low IR optical depth (without requiring radiation trapping); whereas higher values of the momentum ratio ($\zeta \sim 10$) can be achieved by considering large IR optical depths. More precisely, the maximal values of the momentum ratio are reached at small radii, where the outflow is optically thick to the reprocessed IR radiation. As $\zeta_\mathrm{IR} \sim 2 \sqrt{\tau_\mathrm{IR,0}}$, a large momentum boost requires a large IR optical depth at launch. On the other hand, a decay in the central luminosity output may help explain the large values of the momentum ratio observed at large radii. For instance, a power-law decay in luminosity may account for the high observed values ($\zeta \sim 10$) out to large radii ($r \sim 10$ kpc). 

A similar picture is found in the case of the outflow energy ratios, which also span a broad range, with typical values lying between $\sim 10^{-4}$ and $\sim 10^{-2}$ (Fig. \ref{plot_energetics_sample}, right panel). The global distribution may be reproduced by considering different physical conditions in the sources, and in particular the IR optical depth. Since $\epsilon_\mathrm{k,IR} \sim \sqrt{\tau_\mathrm{IR,0}} \frac{v_\mathrm{IR}}{c}$, large values of the energy ratio require large initial IR optical depths, and the maximal values are again obtained at inner radii. But in contrast to the case of momentum boosts, the large $\epsilon_k$-values observed at large radii can be mostly accounted for by just assuming large IR optical depth, without the need to invoke AGN luminosity decay. 
This could be related to the fact that the energy ratio gets an even greater boost than the momentum ratio for enhanced dust-to-gas ratios (due to a steeper dependence on $f_\mathrm{dg}$). 

Overall, the broad range of momentum ratios and energy ratios observed in galactic molecular outflows can be reproduced by considering different physical conditions in the radiation pressure-driven outflows. 
Lower values of the outflow energetics are naturally obtained in the radiative feedback scenario, while higher outflow energetics may be accounted for by including radiation trapping, possibly coupled with AGN luminosity decay \citep[][see also \citet{Zubovas_2018}]{Ishibashi_Fabian_2018}. We note that objects with exceptionally high values of the outflow energetics ($\zeta \gg 10$ and $\epsilon_k \gg 0.05$) are prime candidates for `fossil' outflows (as identified in \citet{Fluetsch_et_2019}). Such fossil outflows are observed in sources with (current) low Eddington ratios, suggesting that they were likely powered by a past AGN outburst that has since faded. 


\section{Outflow scaling relations}
\label{Section_Scaling_Relations}

\begin{figure*}
\begin{multicols}{2}
    \includegraphics[width=0.8\linewidth]{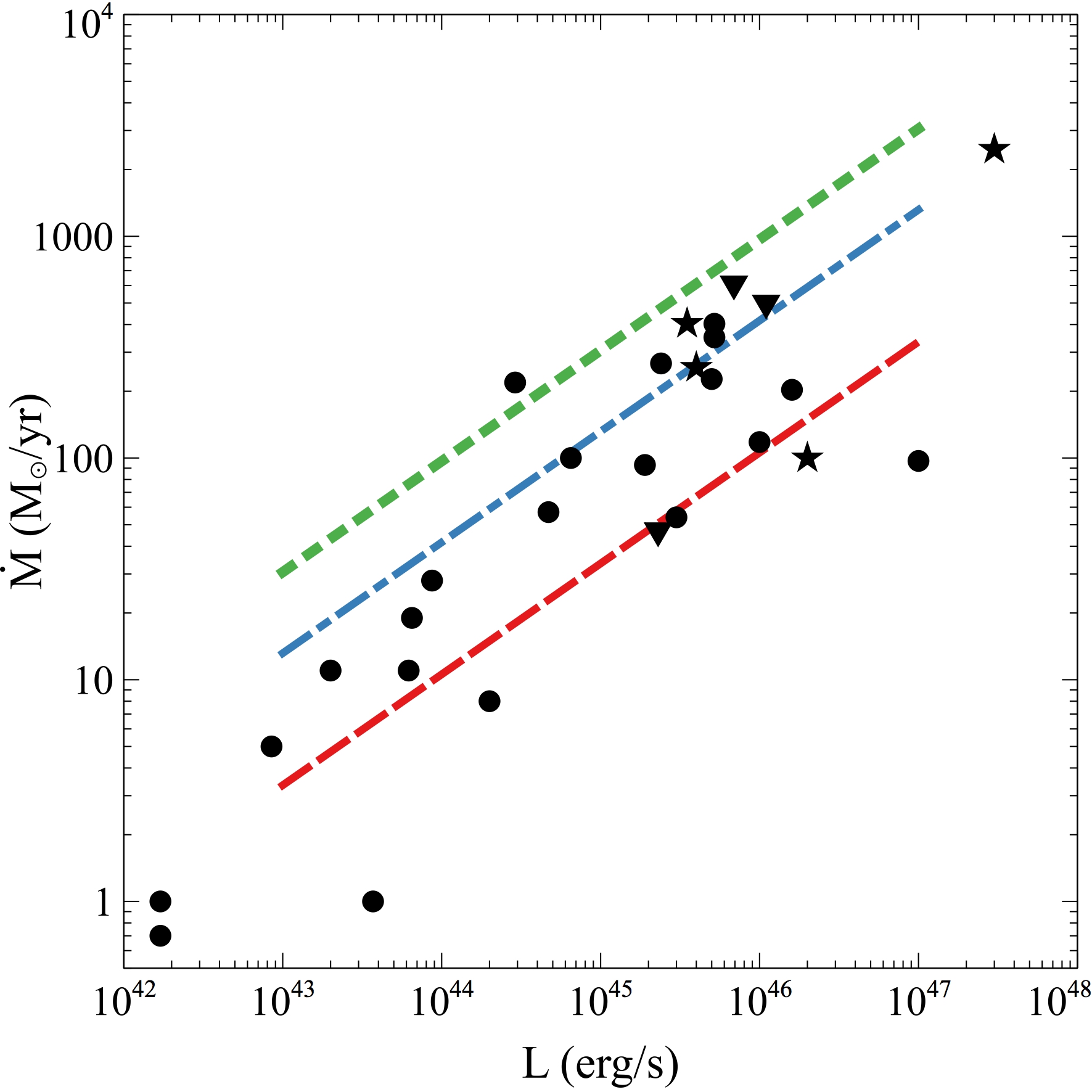}\par
    \includegraphics[width=0.8\linewidth]{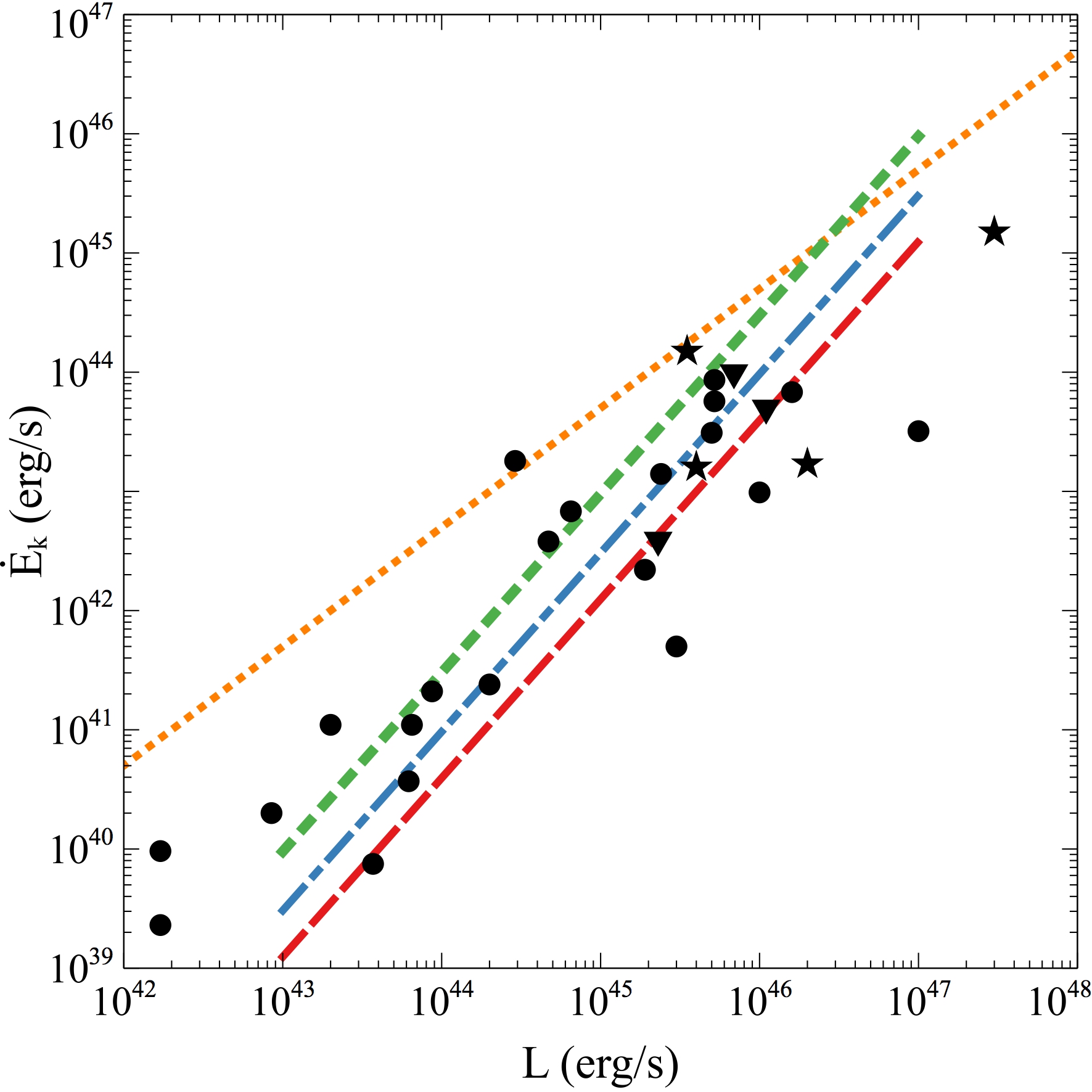}\par 
    \end{multicols}
\caption{ Luminosity scaling relations of radiation pressure-driven outflow models ($r = 1$ kpc and $R_0 = 50$ pc) compared to observational measurements. Model cases: $M_\mathrm{sh} = 10^8 M_{\odot}$, $f_\mathrm{dg} = 1 \times f_\mathrm{dg,MW}$ (red dashed); $M_\mathrm{sh} = 5 \times 10^8 M_{\odot}$, $f_\mathrm{dg} = 3 \times f_\mathrm{dg,MW}$ (blue dash-dot); $M_\mathrm{sh} = 10^9 M_{\odot}$, $f_\mathrm{dg} = 5 \times f_\mathrm{dg,MW}$ (green dotted). 
The orange fine-dotted line marks the standard prediction ($\dot{E}_k/L = 0.05$) of the wind energy-driven model for fiducial parameters. Observational samples: local molecular outflows (dots), upper limits (triangles), additional high-z sources (stars). 
} 
\label{plot_scalings_sample}
\end{figure*}

Correlations between the global outflow properties and the central AGN have been observed in different samples \citep[][and references therein]{Sturm_et_2011, Veilleux_et_2013, Cicone_et_2014, Fiore_et_2017, Fluetsch_et_2019, Veilleux_et_2020}. In general, the outflow rates and kinetic powers are observed to scale with AGN luminosity; in fact, luminosity correlations are to be expected in AGN feedback-driven models. 

Figure \ref{plot_scalings_sample} shows the mass outflow rate and the kinetic power as a function of AGN luminosity for the observational sample described in Section \ref{Section_Global_Comparison} (black symbols). In the same figure, we also plot the predicted analytic scalings derived in Equations \ref{Eq_Mdot}-\ref{Eq_Edotk} (coloured lines). The model scaling relations are computed for a given radius ($r = 1$ kpc), while varying the outflowing shell mass and dust-to-gas ratio (see the figure caption for numerical values). We recall that in our radiative feedback scenario, the mass outflow rate scales with luminosity as $\dot{M} \propto L^{1/2}$, while the kinetic power scales as $\dot{E}_k \propto L^{3/2}$ (Section \ref{Section_Model}). 
 
Different observational scaling relations for molecular outflows have been considered in the literature. A sub-linear scaling of the form $\dot{M} \propto L^{0.76\pm0.06}$ is reported in \citet{Fiore_et_2017}, while \citet{Fluetsch_et_2019} obtain a shallower correlation of the form $\dot{M} \propto L^{0.68\pm0.1}$, based on an enlarged and less biased sample. A flatter slope of $\sim 0.45 \pm 0.04$ is found in X-ray observations of a sample of local galaxies hosting molecular outflows \citep{Laha_et_2018}. Concerning the kinetic power of molecular outflows, a super-linear relation of the form $\dot{E}_k \propto L^{1.27\pm0.04}$ is instead reported \citep{Fiore_et_2017}. Indeed, the $\dot{E}_k/L$ ratio increases with increasing AGN luminosity, and the observed steeper-than-linear relation is found to be consistent in terms of slope with the prediction of the radiation pressure-driven scenario \citep{Fluetsch_et_2019}. On the other hand, as noted in the latter paper, most of the sources tend to fall below the canonical $5 \%$ line predicted by the fiducial wind energy-driven model (but also see Section \ref{Section_Discussion}).  

In Figure \ref{plot_scalings_sample}, we note that there are a couple of objects at the lowest luminosities, which seem to be offset from the main scaling relations. The low values of the present-day luminosity ($L \sim 10^{42}$ erg/s) in these two sources are likely not sufficient to power the currently observed outflows. In fact, a minimal critical luminosity needs to be exceeded to launch a galactic-scale outflow (Section \ref{Section_Model}). It is possible that the outflows we observe today were in fact launched by a powerful AGN episode in the past. If the luminosities were actually higher, the source positions in Figure \ref{plot_scalings_sample} could be shifted towards the right, possibly bringing them somewhat closer to the scaling relations for the kinetic power. 
In principle, we should then also consider the past higher luminosities in the estimates of the momentum and energy ratios (as the present $\zeta$ and $\epsilon_k$ values could be overestimated). 

Most recently, outflowing molecular gas has been observed for the first time in the Milky Way galaxy \citep{DiTeodoro_et_2020}. The detection of this fast cold molecular gas presents a challenge to outflow driving models, since the current level of nuclear activity of SgrA* is clearly inadequate. In the radiative feedback scenario, we estimate that an outflowing mass of $\sim 500 M_{\odot}$, launched from the inner few pc region, would require a central luminosity of the order of $\sim 10^{42}-10^{43}$ erg/s (for a Milky Way dust-to-gas ratio). These values are several orders of magnitude larger than the current luminosity output of the Galactic Centre, and would imply accretion at a $\lesssim$few percent of the Eddington limit. Such a possibility is actually supported by observational evidence, indicating that the Galactic Centre underwent much more active episodes (with bright flares) in the past. For instance, iron line fluorescence measurements indicate that the luminosity of SgrA* was $\mathrm{L_X} \gtrsim 10^{39}$ erg/s in the past few hundred years \citep[][and references therein]{Koyama_2018}. Even greater Seyfert-like luminosities ($\sim 10^{42} - 10^{44}$ erg/s) may be expected a few million years ago, e.g. to explain the Fermi bubbles \citep[][and references therein]{Veilleux_et_2020}. 


\section{UFO+molecular outflow sources in the $\mathrm{N_H} - \lambda$ plane}
\label{Section_NH_lambda}

In a limited number of sources (currently a dozen objects), both a small-scale UFO and a galactic-scale molecular outflow have been detected. UFOs are detected as blueshifted iron absorption lines in the X-ray spectra, while molecular outflows are observed at sub-mm/far-IR wavelengths. It is often assumed that a fraction of the kinetic energy of the inner UFO is transferred to the large-scale molecular outflow on galactic scales. The associated energy transfer rate or coupling efficiency factor $f$ can vary between $0$ and $1$, corresponding to the momentum-conserving and energy-driving limits respectively \citep{Mizumoto_et_2019, Smith_et_2019}.    

Comparison of the momentum rate vs. outflow velocity indicate that the measured momentum boosts can be significantly lower than the values expected from the wind energy-conserving mechanism. For instance, \citet{Reeves_Braito_2019} note that only Mrk 231 and IRAS 17020+4544 (which is a radio-loud source) are consistent with a purely energy-conserving shock. In contrast, in the cases of I Zw 1, PDS 456, and MCG-03-58-007, the observed momentum rates lie nearly two orders of magnitude below the predicted relation in the energy-conserving scenario. In these sources, the observed energetics might be more consistent with momentum-driven outflows \citep{Reeves_Braito_2019, Bischetti_et_2019, Sirressi_et_2019}. More in general, some objects may belong to the momentum-driven regime, while others may be closer to the energy-driven mode, depending on the specific physical conditions in each particular source \citep{Smith_et_2019, Chartas_et_2020}.  

This suggests a wide range of coupling efficiencies $f \sim 0.001 - 0.5$ (with a preference for lower values) relating the inner wind and the outer molecular outflow \citep{Smith_et_2019, Mizumoto_et_2019}. If interpreted in terms of shocked winds, the observed low energetics would require a rather weak and inefficient coupling between the nuclear UFO and the interstellar medium on galactic scales. Alternatively, the galactic outflows could be driven by other physical mechanisms, such as radiation pressure on dust, which naturally lead to low values of the momentum and energy ratios.

\begin{figure}
\begin{center}
\includegraphics[angle=0,width=0.4\textwidth]{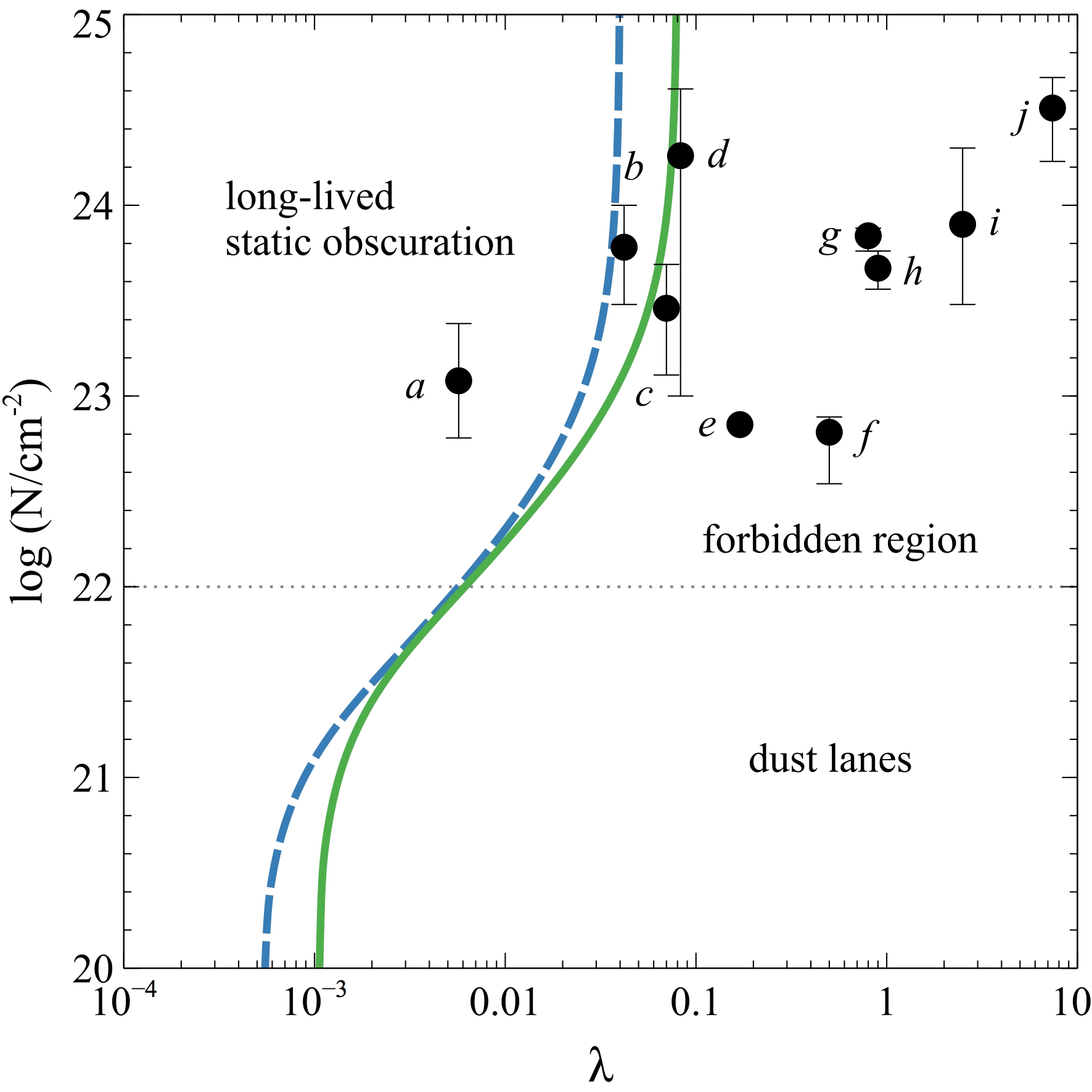} 
\caption{ 
Location of sources with both UFO and molecular outflow detections in the $\mathrm{N_H} - \lambda$ plane. The observational data points are retrieved from different works in the literature and labelled in order of increasing $\lambda$ (as listed in the main text). 
The model curves represent two cases of differing dust opacities \citep[cf.][]{Ishibashi_et_2018b}: $\kappa_\mathrm{IR} = 5 \, \mathrm{cm^2 g^{-1}}$ and $\kappa_\mathrm{UV} = 400 \, \mathrm{cm^2 g^{-1}}$ (green solid), and $\kappa_\mathrm{IR} = 10 \, \mathrm{cm^2 g^{-1}}$ and $\kappa_\mathrm{UV} = 800 \, \mathrm{cm^2 g^{-1}}$ (blue dashed). The horizontal line at $\mathrm{log \, N} = 22$ (grey fine-dotted) marks the limit below which absorption by outer dust lanes becomes dominant. 
}
\label{Fig_NH_lambda}
\end{center}
\end{figure}

The importance of radiation pressure on dust can be constrained by analysing the location of the sources in the $\mathrm{N_H} - \lambda$ plane,  defined by the column density versus the Eddington ratio \citep[][]{Fabian_et_2008, Fabian_et_2009, Ishibashi_et_2018b}. The effective Eddington limit for dusty gas delineates two distinct areas in this plane: the region of `long-lived obscuration' (to the left of the critical curve) and the so-called `forbidden region' (to the right of the dividing line). In Figure \ref{Fig_NH_lambda}, we show the location of ten sources having both UFO and molecular outflow detections (along with X-ray column density estimates) in the $\mathrm{N_H} - \lambda$ plane. The objects are collected from the literature and are listed in order of increasing Eddington ratio (with corresponding labels in brackets): IC 5063 [\textit{a}] \citep{Mizumoto_et_2019}, NGC 1068 [\textit{b}] \citep{Mizumoto_et_2019}, HS 0810+2554 [\textit{c}] \citep{Chartas_et_2020}, NGC 6240 [\textit{d}] \citep{Mizumoto_et_2019}, IRAS F05189-2524 [\textit{e}] \citep{Smith_et_2019}, APM 08279+5255 [\textit{f}] \citep{Feruglio_et_2017, Hagino_et_2017}, PDS 456 [\textit{g}] \citep{Nardini_et_2015, Bischetti_et_2019}, I Zw 1 [\textit{h}] \citep{Reeves_Braito_2019}, Mrk 231 [\textit{i}] \citep{Mizumoto_et_2019}, F11119+3257 [\textit{j}] \citep{Veilleux_et_2017, Tombesi_et_2017}. We observe that most of the sources (9 out of 10) are located in the forbidden region (or close to borderline) where outflows are expected; and only one source (which is a radio-loud AGN) lies in the region of long-lived obscuration. This reinforces the notion that radiation pressure on dust may play an important role in driving the observed galactic outflows.

\section{Discussion}
\label{Section_Discussion}

Comparison of the AGN radiation pressure-driven model with observations of galactic molecular outflows shows that the global distribution of momentum ratios and energy ratios can be plausibly accounted for: low values of the outflow energetics are a natural outcome of radiative feedback, whereas higher energetics may be reproduced by invoking IR radiation trapping and/or luminosity decay. Furthermore, the AGN radiative feedback scenario predicts luminosity scalings (a sub-linear scaling for the mass outflow rate and a super-linear scaling for the kinetic power) quantitatively consistent with the observational scaling relations. We thus find a nice overall agreement between model predictions and observations over a wide range of AGN and host galaxy parameters. 

\citet{Menci_et_2019} consider the wind shock model in a disc geometry and perform a detailed comparison with molecular outflows in single objects, as well as a broader comparison with a large sample of (mostly) ionised outflows. Focusing on the radial profiles of the outflow velocity and mass outflow rate, they also analyse the outflow scaling relations with AGN luminosity. In all such comparisons, one should keep in mind that the model predictions are computed at a particular radius $r$, while the observational values are distributed over a range of radial distances. Two radial bins, divided into small-scales ($r < 1$ kpc) and large scales ($r > 1$ kpc), are thus considered in \citet{Menci_et_2019}. Given that, on average, molecular outflows tend to be located at smaller radii compared to ionised counterparts, our choice of $r = 1$ kpc for the computation of the model energetics in Section \ref{Section_Scaling_Relations} should be acceptable. 
In any case, accurate radial and angular dependences of the outflow properties can not be achieved by currently available observations. (This is also why \citet{Menci_et_2019} do not make use of the full power of their two-dimensional analysis). Future observations with higher resolution and improved sensitivity should allow us to better disentangle between different model predictions.  
 
An empirical constraint on the relative importance of radiation pressure versus hot gas pressure may be obtained by the analysis of emission line ratios. The observed line ratios in the average quasar spectrum are consistent with a radiation pressure-dominated scenario over a wide range of radial distances ($r \sim 0.1 \, \rm{pc} - 10 \, \rm{kpc}$) \citep{Stern_et_2016}. Spatially resolved UV emission spectroscopy of the outflow in the obscured quasar SDSS J1356+1026 indicates that the UV line ratios are very similar in the nuclear region ($r \lesssim 100$ pc) and on galactic scales ($r \sim 10$ kpc) \citep{Somalwar_et_2020}. These observations constrain the ratio between hot gas pressure and radiation pressure to $P_\mathrm{hot}/P_\mathrm{rad} \lesssim 0.25$ at all radial scales, suggesting that radiation pressure is likely at the origin of the observed outflowing gas in SDSS J1356+1026. 

In our AGN radiative feedback scenario, there are two important conditions for efficient outflow driving: radiation trapping and dust content. 
From the observational perspective, the large IR optical depths required for radiative feedback may be reached in the nuclear regions of obscured ULIRG-like systems. Recent sub-mm and far-IR observations reveal that some ULIRGs host extremely compact obscured nuclei, characterised by huge column densities (up to $\mathrm{N_H} \sim 10^{26} \mathrm{cm^{-2}}$) and Compton-thick obscuration on tens of pc scales \citep[][and references therein]{Aalto_et_2019b}. Such heavily obscured and buried nuclei can be optically thick to the IR and even sub-mm wavelengths. For instance, ALMA observations of the nearby infrared luminous galaxy IC 860 indicate that the mm-continuum emission is dominated by dust, leading to significant dust opacities at mm-wavelengths \citep{Aalto_et_2019a}.  

Another important requirement in the radiative feedback scenario is the presence of significant amounts of dust, which sustain the overall AGN feedback process. Dust grains can be produced in large quantities in core-collapse supernovae \citep{Owen_Barlow_2015, Wesson_et_2015}, enriching the local environment in nuclear starbursts. Moreover, grain growth in the interstellar medium can also contribute to the increase of the dust mass at later times \citep{Michalowski_2015}, while turbulence may further accelerate the growth of dust grains \citep{Mattsson_2020}. The associated dust-to-gas ratios can be quite high, with values up to $\sim 1/30$ in dust-reddened quasars \citep{Banerji_et_2017}, and potentially up to $\sim 1/20-1/10$ in heavily dust-obscured ULIRG nuclei \citep{Gowardhan_et_2018}. The combination of high column densities and substantial dust content should lead to large IR optical depths, forming particularly favourable conditions for AGN radiative feedback. 

On the theoretical side, recent radiation hydrodynamic (RHD) simulations of radiation-driven shells indicate that the boost factor is roughly equal to the IR optical depth ($\dot{p} \sim \tau_\mathrm{IR} L/c$ for moderate optical depths), confirming the simple analytic picture \citep{Costa_et_2018}. The correspondence may break down at the highest IR optical depths ($\tau_\mathrm{IR} \sim 100$), whereby the photon diffusion time becomes longer than the flow time, and the relation becomes sub-linear. From our analytic scalings, we expect the maximal value of the momentum ratio to scale as the square root of the initial IR optical depth ($\zeta_\mathrm{IR} \sim \sqrt{\tau_\mathrm{IR,0}}$). The importance of radiation trapping is further corroborated in new RHD simulations of AGN radiative feedback, which report an excellent agreement between the simulated outflow propagation and the expected analytic solutions, when including IR multi-scattering \citep{Barnes_et_2020}. 

Interestingly, most of the sources with both UFO and molecular outflow detections are located in the forbidden region of the $\mathrm{N_H} - \lambda$ plane (Section \ref{Section_NH_lambda}). Similarly, highly dust-reddened quasars at high redshifts ($z \sim 2$), as well as other populations of luminous reddened quasars, are observed to lie in the forbidden region, suggesting that they are indeed in a blowout phase \citep{Lansbury_et_2020}. On the other hand, dust obscured galaxies (DOG) may be located just at the boundary between the forbidden and long-lived regions \citep{Zou_et_2020}, and could potentially represent sources at a slightly earlier evolutionary phase. All these observations support the important role of radiation pressure on dust in powering galactic outflows. 

Finally, we remark that AGN radiative feedback naturally fits into the `co-evolutionary sequence' relating dust-obscured starbursts to unobscured luminous quasars \citep[e.g.][]{Sanders_et_1988}. In most co-evolution scenarios, some form of AGN feedback is required to blow out the obscuring gas and dust, but the actual mechanism remains often unspecified. Radiation pressure directly acts on the obscuring dusty gas, with the more dusty gas being preferentially removed by radiative feedback \citep{Ishibashi_Fabian_2016b}. 
In our picture, AGN radiative feedback can thus provide a natural physical interpretation for the co-evolutionary path. 


\section{Conclusion}
\label{Section_Conclusion}

Despite the huge observational progress in the detection and characterisation of galactic outflows, the underlying driving mechanism is still ill-defined. Here we directly compare the AGN radiative feedback model with the most recent compilation of galactic molecular outflows. We show that radiation pressure on dust can quantitatively account for the dynamics and energetics of galactic outflows (Section \ref{Section_Comparison}), as well as the observational scaling relations (Section \ref{Section_Scaling_Relations}). The agreement we find between model predictions and observations, over a wide range of AGN and host galaxy parameters, is quite encouraging. The simple analytic picture of radiation feedback on dusty gas also seems to be broadly supported by the latest RHD numerical simulations. Therefore we should move towards a consensus acknowledging the role of radiation pressure on dust in driving galactic outflows, alongside wind-driving. Indeed, both wind feedback and radiation feedback need to be considered in the interpretation of future observational data to better constrain the physical origin of galactic outflows. 


\section*{Data availability}

No new data were generated or analysed in support of this research.


\bibliographystyle{mn2e}
\bibliography{biblio.bib}

\label{lastpage}

\end{document}